# Human anaerobic intestinal "rope" parasites

Alex A. Volinsky [a*], Nikolai V. Gubarev [b], Galina M. Orlovskaya [c], Elena V. Marchenko [d]

[a] Department of Mechanical Engineering, University of South Florida, Tampa FL 33620, USA

[b] Occupational Safety Ltd. (OOO "Bezopasnost Truda"), 32 Koli Tomchaka St., suite 14, St. Petersburg 196084, Russia

[c] Department of Surgery, St. Petersburg City Hospital No. 15, 4 Avangard St., St. Petersburg 198205, Russia

[d] Currently no affiliation, Formerly research volunteer at H. Lee Moffitt Cancer Center and Research Institute, 12902 USF Magnolia Drive, Tampa FL 33612, USA

[*] Corresponding author. Phone: +1 813 974 5658, Fax: +1 813 974 3539, Email: volinsky@usf.edu

**Abstract**

Human intestinal helminths are described in this paper. They can be over a meter long, with an irregular cylindrical shape, resembling a rope. These anaerobic intestinal "rope" parasites differ significantly from other well-known intestinal parasites. Rope parasites can leave human body with enemas, and are often mistaken for intestinal lining, feces, or decayed remains of other parasites. Rope parasites can attach to intestinal walls with suction bubbles, which later develop into suction heads. Walls of the rope parasites consist of scale-like cells forming multiple branched channels along the parasite's length. Rope parasites can move by jet propulsion, passing gas bubbles through these channels. Currently known antihelminthic methods include special enemas. Most humans are likely hosting these helminths.

**Keywords:** Human parasite; Intestinal parasite; Anaerobic parasite; Helminths; Rope parasite; New taxa.

**Introduction**

More than one billion humans are infected with intestinal parasites.[1] Over 15% of all cancers in humans are parasites-related.[2] It is also estimated that every fourth human is infected with intestinal parasites.[3] Parasitic worms fall under one of the four categories: roundworms (nematodes), tapeworms (cestodes), flukes (trematodes), and monogeneans.[4]

This communication describes human anaerobic intestinal parasites that have not been previously mentioned in the literature. In humans they can cause multiple symptoms, including weight gain or loss, food allergies, common colds, coughing, back pain, rashes, headaches, indigestion, hair loss, etc. Discovered anaerobic intestinal parasites differ significantly from the well-known and studied intestinal parasites.

Unlike others, these parasites do not have muscles, nervous system, or distinct reproductive organs, etc., and dry out quickly when exposed to air. The main reason these parasites have not been previously discovered by the researchers, is because they rarely come out as whole fully developed adult species. They also look like human excrements (Fig. 1(a)), and don't move outside the human body in air. These parasites are often mistaken for decaying remains of other parasites, feces, or lining of the intestines, as their colour varies from white to brown, to dark grey. Fig. 1(b) shows over a meter long adult intestinal parasites, which resembles a rope. Thus, the adult stage of these parasites is named "rope" parasites. Rope parasites can leave human body with special enemas,[5-8] which is how they have been discovered.

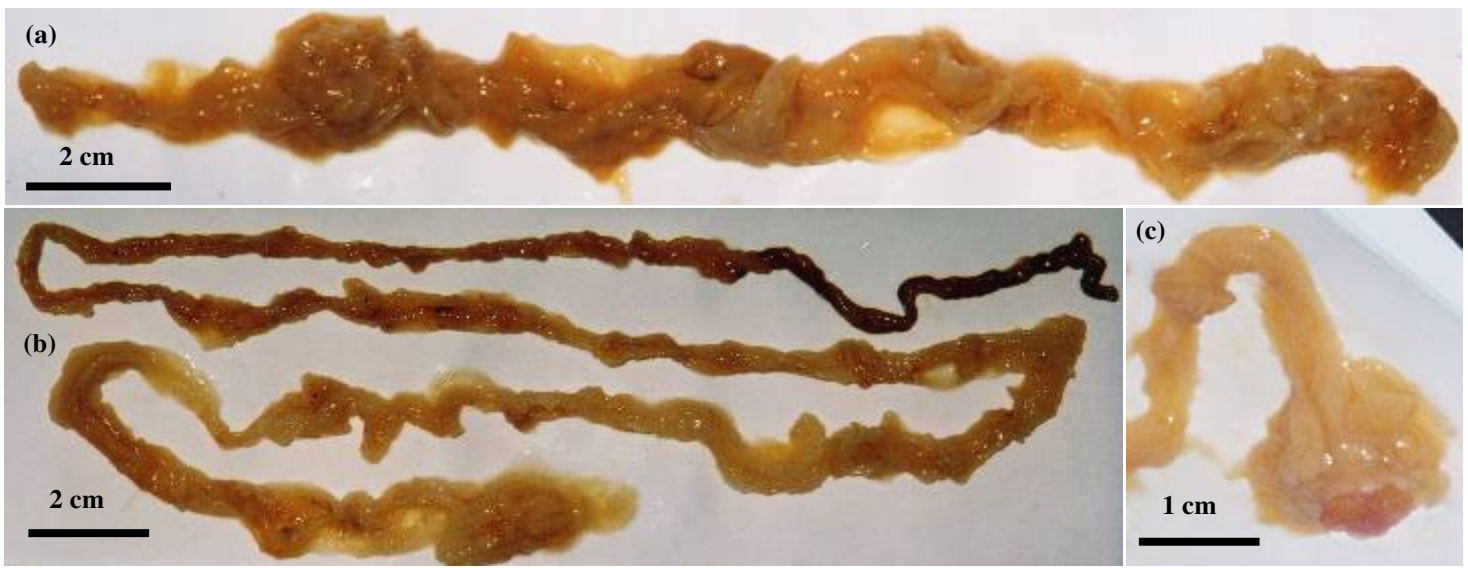

Fig. 1. (a) Anaerobic human intestinal "rope" parasite; (b) Fully developed adult intestinal parasite with the suction head; (c) Close-up view of the suction cup head.

**Results**

Adult rope parasites have an un-segmented irregular cylindrical shape. Parasites' tegument is slimy and tacky. Once washed with water from feces, rope parasites produce a very strong distinct scent. Microscopic observations of the adult parasite structure are presented in Fig. 2. Fig. 2(a) is an optical micrograph of the parasite's wall, which has multiple branching channels, 50-150 µm wide. The walls are formed by multiple scale-like cells, 10-20 µm in size, clearly seen in Fig. 2(a) and (b). When water was injected into the channel, it travelled along the length of the parasite, observed under the optical microscope, without carrying any surrounding cells. There was also some particles motion in the channels, observed under the optical microscope. Unlike nematodes, which have the main body cavity, rope parasites have a branching network of channels along their length (Fig. 2(a). Thus, unlike other nematodes, they have no single apparent digestive tube. Similar to some nematodes, rope parasites attach to the internal wall of the intestine using suction cups positioned at the head (Fig. 1(c)), or even along their body (Fig. 1(a)). At the same time, these parasites don't have a cuticle, like nematodes, but a tegument, like cestodes. This tegument is slimy and gel-like in consistency. Rope parasites are most likely hermaphroditic, as no reproductive organs were found by either microscopic or macroscopic observations. They have no intermediate host and no apparent larval stages.

Fig. 2(c) shows almost perfect spherical objects, 35-40 µm in diameter found inside the channels. Larger magnification image of this feature is shown in Fig. 2(d). Such spheres were observed inside the two parasites from two hosts subjected to microscopic examination. While at this point it's not clear what these spherical objects inside the channels are, it is quite possible that they are simply gas bubbles. Fig. 2(e) shows slime originating from the early stages of the rope parasite development. Notice blunt droplet ends, which later develop into attachment suction cups. Fig. 2(f) shows rope parasite placed in water. Originally they were laying down on the bottom of the glass jar. But later, rope parasites developed much larger gas bubbles, attempting to reach water surface, defying gravity. This behaviour points to their motion and attachment mechanisms. It is quite possible that parasites combined smaller bubbles in Fig. 2(d) into larger bubbles seen in Fig. 2(f). Rope parasites can also emit gas and move by jet propulsion. This process is manifested in humans by flatulence and stomach bloating during or after enemas. Rope parasites utilize osmotic feeding, as food remains were observed inside them, determining their colour, as in Fig. 1(a) for example. In the intestines adult parasites expand, while twisting like a corkscrew, or a mop, capturing fecal content and extracting nutrients, which produces dry twisted feces. By doing this, rope parasites can completely block the lumen of the intestine. Feeding and excretory functions are conducted through their whole body surface. Two possible reproductive routes through gemmation and slime produced by the rope parasites can be named.

**Discussion**

A legitimate question how these parasites can stay inside the human intestinal tract without being carried out by peristaltic movements with excrements arises. First, rope parasites attach to intestinal walls with suction cups. Second, they are quite long, reaching over a meter in length, exceeding a typical length of the fecal contents. Third, they move by emitting bubbles seen in Fig. 2(c), utilizing jet propulsion. Fourth, they twist like a mop, expanding in cross-section, completely blocking the lumen of the intestine. Fifth, they develop gas bubbles, as in Fig. 2(f), which later become suction heads, as in Fig. 1(c). These mechanisms allow rope parasites to resist peristaltic waves, which only travel a few centimetres per second. Rope parasites also bypass human immune system and participate in the digestion process by consuming fecal contents, striping humans of vital nutrients, and generating toxins as their own waste in return. They are most active during the night time.[8] The exact mechanisms are yet to be discovered and studied.

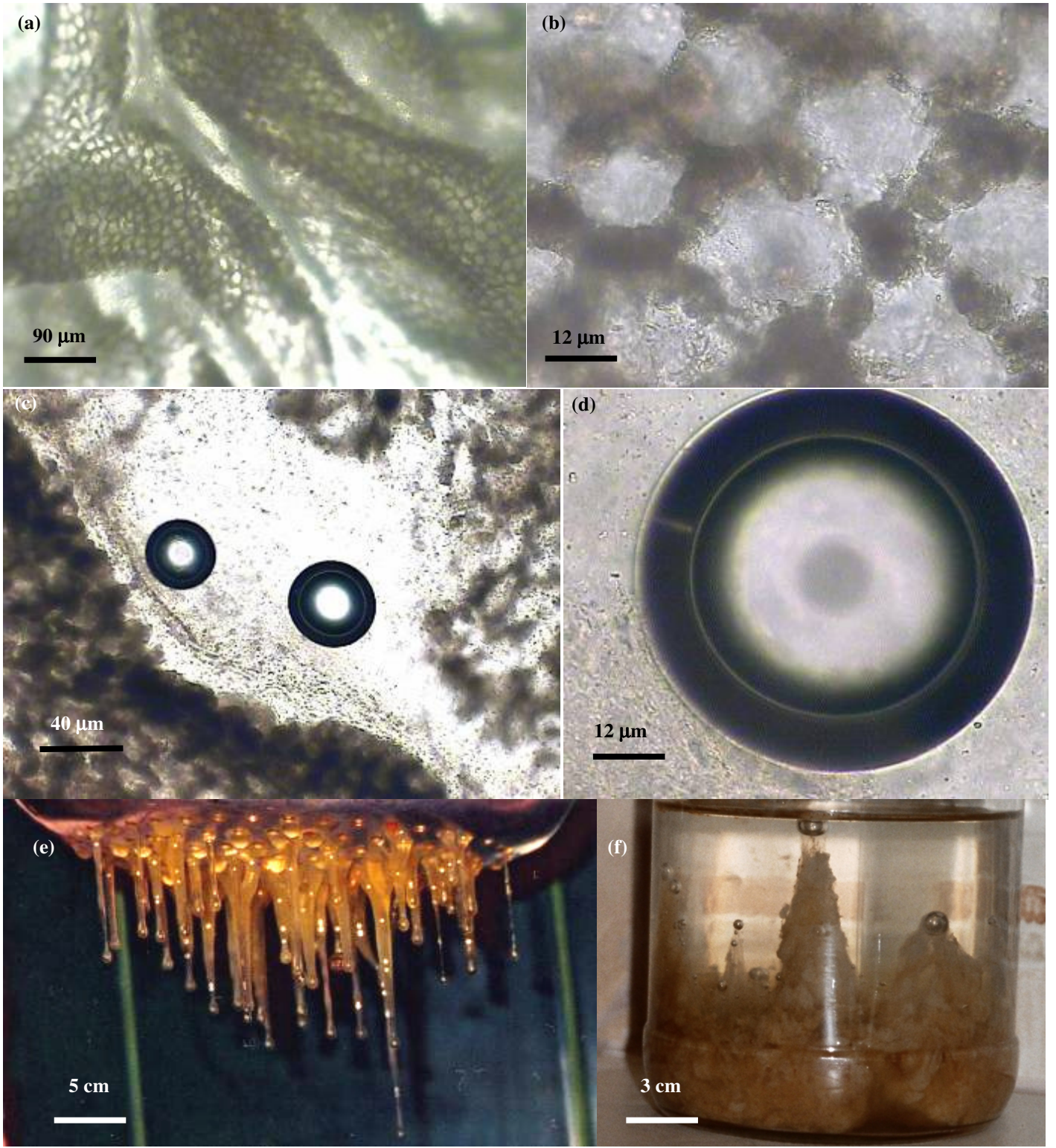

Fig. 2. (a) Optical micrograph of the rope parasite wall structure showing scale cells and microchannels. Unstained sample; (b) Higher magnification optical image of the parasite wall; (c) Spheres inside the channel cavity; (d) Higher magnification image of a sphere inside the rope parasite channel cavity; (e) Slime initial stages of the rope parasite' development coming from the colander holes; (f) Rope parasite in water generating gas bubbles rising to the surface.

Many people observe rope parasites in their stool during fasting or enemas, with numerous pictures posted in corresponding internet forums worldwide. Often the rope parasites are mistaken for the remains of other decayed parasites, such as ascaris worms, or the intestinal lining. It is quite possible that fully developed rope parasites can feed on blood, as some of them came out after enemas with their head covered with fresh human blood. People with alkaline blood reaction (blood pH of 8-10) had the worst parasitic invasions, most likely due to chronic constipation caused by intestinal dysbacteriosis, which rope parasites contribute to significantly. An easy way to determine if blood has alkaline reaction is based on the eye conjunctiva colour behind the eyelid. Bright pink conjunctiva colour signifies normal blood pH, while bright red colour corresponds to alkaline blood, and pale conjunctiva corresponds to acidic blood. Similar to Giardia,[9] rope parasites release toxins in the intestinal tract, and the blood stream, suppressing immune system and causing multiple symptoms in humans. Once an adult parasite is attached to the intestinal wall, human body does not have mechanisms to get rid of it. Rope parasite tegument releases a special slimy acrid substance with a distinct smell. This substance helps parasites attach to the lining of the intestines and protects them from proteolytic enzymes. It is quite possible that this slime protects the parasites from human body immunological response. This slime release is seasonal, and it can travel up the intestinal tract, reaching lungs and airways. One can only imagine how many microelements, vitamins and enzymes the human body is stripped off to sustain this parasitic biomass. Human body starts to lack essential nutrients, which leads to higher food consumption, resulting in obesity. These helminths themselves also contribute to the excess body weight.

While currently there's no known pharmaceutical antihelminthics cure for the rope parasites, dehilminthation methods include cleaning water enemas, followed by enema with baking soda,[5,8] eucalyptus oil followed by the lemon juice enema,[6,7,8] and milk enemas with salt.[8] Most humans are hosting these parasites without even knowing it,[8] as hundreds of patients had these parasites. So far there has not been a single patient that didn't have these parasites. The rope parasite's full life cycle and interactions with the human body are yet to be studied and fully understood.

**Conclusions**

In conclusion, human anaerobic intestinal parasites, named "rope" parasites, have been described. They feed osmotically on semi-digested food remains in the intestines, and move by utilizing jet propulsion by releasing gas bubbles. Rope parasites attach to the intestinal wall and cause multiple symptoms. Rope parasites can leave the human body with enemas.